# THE QUEST FOR DEVELOPMENT: WHEN SOCIAL MEDIA-BROKERED POLITICAL POWER ENCOUNTERS POLITICAL 'FLAK JACKETS'

Boluwatife Ajibola, London School of Economics and Political Science (LSE), ajibola.boluwatife50@gmail.com

**Abstract:** Social media provides an extended space for collective action, as netizens leverage it as a tool for claim-making and for demanding the dividends of governance. However, political regimes often greet expanding use of social media with censorship, which netizens often have to contend with, particularly in the quest for development outcomes. While existing studies having expansively explored multiple uses of social media, the specific features that signal their massive uptake and how this intersects with the quest for political power has not been substantially documented. This paper argues that social media is characterized by social buttons that expedite the multiplication of 'digital bullets' – in forms of tweets and perceived detestable comments – which compromise the defense lines of political regimes, hence, their uptake of censorship as metaphorical 'flak jackets'. This research is conducted on the basis of key informant interviews with voices against social media censorship in Nigeria since the inception of Nigeria's ruling government in 2015, particularly following the proposed 'Protection from Internet Falsehood and Manipulations Bill' in 2019.

## 1. INTRODUCTION

Social media has not only metamorphosed communication lines between and among the peoples of the world, easing interactions and redefining the complexities of human association, but its uptake in civic engagement has also expanded in the last two decades. Pre-existing mediums of information exchange and communication has shifted, to adapt to the presence of social media. As a source for news about public matters (Newman et al. 2014), and conduit for citizens' political engagement (Boulianne, 2019), social media has connected the world and eliminated boundaries (Van Dijk, 2006). The meagre cost of content exchange among connected audiences, particularly in creating public awareness partly explains its popularity as a tool for political engagement. Importantly, some social networks possess unique features through which audiences gain access to contents.

The instrumentality of social media, and the internet in general, in social networking have undoubtedly transformed the political landscape as people have moved largely from being passive audiences to political messages (Evans, 2010). Political office holders often leverage social media networks in projecting their manifestos, personal and party ideologies; in gaining visibility; reporting periodical achievements; and some in seeking public or sectional approval. However, some, through their social media accounts, are often criticized by aggrieved citizens. Some regimes present themselves as democratic, having pledged their commitment to the promotion of its tenets, however, at some point the feedback lines, or the norm of engagement becomes unacceptable to them, causing them to act in ways contrary to democratic principles.

Vareba et al (2018) notes that internet regulation has become an indispensable tool today for the perpetuation of varied forms of vices, and governmental regimes in both democratic and autocratic settings have implemented some of these to restore some forms of 'sanity'. However, it is found that social media regulations in Africa have been more of threat to freedom than a panacea to checking online vices. African governments have frequently deployed internet regulation as an instrument of political intimidation and violation of human rights, and as a means to undermining the acquisition and expression of political power by the citizens. While several researchers have investigated the positive and negative uses of the social media, as well as the varied forms of its censorship, inviting further investigation are new research into how political censorship of social media, as a result of its affordances, threaten political power and consequent socio-economic development. So far, also, there has been little discussion about the democracy-development nexus, in the context of evolving privileges of social media to citizens, especially in Africa.

Acemoglu and Robinson (2008) theorize that for economic institutions and policies to be changed in a credible way, people need to acquire political power. They distinguish between two components of political power – *de jure* (institutional) and *de facto* political power. D*e jure*, political power originates from political institutions in the society, while *de facto* political power originates from outside the sphere of political institutions, whereby, people through revolts or collective action pursue and impose their development objectives. With *de jure* political power originating from political institutions, one of which is democracy – *"a set of political institutions"* (Acemoglu and Robinson, 2008), and *de facto* leveraging collective action, it is imperative to understand how socio-economic development can be hindered through the repression of digital tools or contraction of spaces through which people acquire and demonstrate political power. Ultimately, this paper argues that social media enhances the acquisition of political power requisite for securing desired socio-economic development objectives including those not prioritized by political regimes. It goes further to conclude that netizens leverage 'digital bullets' of social media in acquiring political power – actions which are in-turn greeted by repression by political regimes, especially when their 'defense lines' are perceived to have been compromised.

This paper has been divided into four main sections. The first section reviews literature on social media uses, their censorship, and political power. This is followed by the section outlining the methodology of primary research, then the presentation of findings on anti-social media legislation popular under the President Muhammadu Buhari administration in Nigeria. These are followed by the discussion section and the conclusion. The findings of this study will contribute a robust and nuanced knowledge to that existing on the development-democracy nexus, and social media censorship in the global South, while it will provide more contextualized depth to those of the North.

## 2.   SOCIAL MEDIA USES, CENSORSHIP AND POLITICAL POWER

Social media has largely diffused content creation processes. As opposed to mainstream media which has to pay journalists for distributable contents, social media platforms such as Facebook, Twitter, YouTube and Instagram provide 'spaces' where contents can be uploaded by the general public and citizenry (de Zwart, 2018). It has also brought the world to an era of more "interactive communication indicated by the emergence of new media", whereby anyone can be a producer of information (McQuail, 2005:40); causing a shift to a more social environment, where users are not just passive receivers, but are also content creators (Bruns, 2008; Jha & Bhardwaj, 2012). This extends from the viewpoints of Bryer & Zavatarro (2001) who revealed that social media facilitates social interaction, makes collaboration possible, and enables stakeholder-deliberation.

Studies have found a positive correlation between online political expression and political participation. The privilege social media grants to freedom of political expression have largely escalated online and offline political engagement (Bode et al, 2014; Hsieh & Li, 2014; Holt et al, 2013; Tang & Lee, 2013), thereby liberalizing democratic spaces. Further, the extent to which people are empowered to participate in political practices, determine the extent of democratic acceptance. Participation here includes voting, representing, deliberating, resisting, and contributing to self and collective rule (Warren 2017). Corroborating this are scholars (Beauvais 2017; Fung 2013; Goodin 2007; Young 2000) who find that inclusion is at the fore of democratic processes, since decisions are only democratic to the degree that those affected by collective outcomes are empowered to influence them.

Further, despite its innumerable positive benefits, social media, as found, aids the creation of "false identities and superficial connections, and is a primary recruiting tool of criminals and terrorists" Amedie (2015). In so far as social networks grant free expression of thoughts, bridges gaps between the government and the governed, and spurs social change, it can be argued that some employ social media as revenge tools against the government in question. While some of these attacks are somewhat constructive and correctional towards desired reparations, they are sometimes perceived as derogatory, thus deserving legal rebuttal. In addition, sanctions, government threats, laws and ordinances are attempted at bridling the uptake of social media in channeling grievances.

Studies on social media and politics in Africa have also explored the privileges of social media to social movements and political regimes. It is admitted that the kinds of media used by social movements are crucial to processes of claim-making (Dawson, 2012), hence social movements prefer mediums that would rather make them less visible to the public or government, as opposed to mainstream media. In their study of social movements in South Africa, Dawson (2012) found that social media deepens discursive opportunities as well as backstage interactions and mobilisations. This notion is attested to by other studies (Vasi et al., 2015; Mausolf, 2017). For example, Nyabola (2018) finds evidence from Kenya as presenting social media platforms as toxic spaces that replicate offline harm. Their use in Kenya encourages and aids the distribution of violent acts. Connected violence are not only between citizens and the government, but politicians are also seen threatening themselves with physical violence on these platforms. In Nigeria, the government has repeatedly blamed social media as a channel for what is referred to as 'fake news' and 'hate speech' (Cheeseman et al., 2020). Tendencies for the spread of 'fake news' and 'hate speech' in Nigeria is blamed on social media's provisions for anonymity and ubiquity (Udanor & Anyanwu, 2019), and this makes combating it difficult.

Governments respond to perceived negative affordances of social media by making attempts to censor their usage by their citizens. This has been severally demonstrated by countries in Africa through sophisticated and deceptive means. Gumede (2016) conceives social media censorship in Africa as attempts by regimes to "silence democratic opposition, civil society and activists' mobilization against poor governance" (Gumede, 2016.p413) through grievances channeled via the platforms. It is further captured that the increasing uptake of social media by Africans is as a result of the continent's history of monopolized mainstream media by political affiliates or politicians, which largely leads to the suppression of dissenting voices or critics who make daring demands on governments (Okocha & Kumar, 2018; Gumede 2010). In answering the question about why governments restrict media freedom, Kellam & Stein (2016) contend that governments are often opposed to ideological stances which do not reflect theirs, and this is common in contexts where the legislature and judiciary hold weak powers relative to the executive. Evidence in Uganda and Zimbabwe (Cheeseman et al., 2020) attests to this. In Nigeria, it is often argued that social media censorship poses more of a threat to freedom of speech than being a panacea to checking online vices (Vareba et al. 2018).

Furthermore, how do we triangulate the repression of dissenting voices and the political and economic institutions foundational to development? Political institutions have fundamental implications on economic institutions (Besley & Kudamatsu, 2006; Mueller, 2007). As noted by Acemoglu & Robinson (2008), democracy being a set of political institutions possess potentials for economic as well as development outcomes. Moreover, the distribution of *de jure* political power can be altered by changes in the political institutions, but to partly or wholly offset changes in *de jure* political power, the institutions create incentives for investments in *de facto* political power. Acemoglu and Robinson (2008) presents this as a model that can imply a pattern of captured democracy whereby the democratic regime may survive, however, the elites reap benefits from the economic institutions. The termination of the Malthusian cycle and the economic growth witnessed after the 17th Century conflicts in Britain resulted in a set of economic institutions which gave property rights to a wide range of people (Thompson 1975). Consequently, the exercise of *de facto* political power was diffused to the poor and politically disenfranchised groups (Tarrow, 2011).

Acemoglu & Robinson, (2005) further revealed that the rise in *de facto* political power of the poor orchestrated the change in political institutions in the favor of those poor and politically disenfranchised groups. This was a foundation to the future allocation of *de jure* political power, it as well served as a means to ensuring that economic policies in the future would be in their interests. However, in contemporary times, political actors are seen to be threatened by power in the hands of the majority, hence, their repressive tendencies (Suter et al. 2005).

With ongoing evidence on social media use, as well as its repression as a result of its perceived negative privileges by political regimes, reviewed above, more needs to be researched into the development implications of the contraction of social media spaces, as well as how these implications can be understood through the lens of *de jure* and *de facto* political power. Through the insights provided by Acemoglu and Robinson's theorization about the relationship between institutions and political power, it can be indicated that social media provides an extended space for people to acquire and demonstrate political power. The political power acquired through political institutions (democracy) – *de jure*, and that acquired through revolt and collective action – *de facto*, provides the latitude for people to make demands for changes and policies that are requisite for socio-economic development in any society. As put by Acemoglu and Robinson (2008), with an increased *de jure* political power, citizens are more able to secure the economic institutions and policies that they so desire. Hence, by their demand for democracy through some forms of collective action, they acquire more 'political say and political power tomorrow' (Acemoglu and Robinson, 2008).

Notably, social media provides an extended space for these forms of collective action, arising from the mobility the platforms afford users between online and offline spaces, essentially, the latitude to either continue their offline actions, or vice-versa, and perhaps grant the privilege for simultaneous multi-cited actions (Allmann, 2016). Political regimes are not oblivious to these realities, hence they attempt to control political institutions, as a means to 'regulating the future allocation of political power' (Acemoglu and Robinson, 2008). Until recently, there has been no substantial analysis of the features of specific social media platforms and how their affordances are perceived by governments, or how they intersect with political power. Hence, what is about the social media that gets political regimes tremulous? Or what about it galvanizes their defensive and belligerent actions? Furthermore, this article contends that social media is a characterization of social buttons (Gerlitz and Helmond 2013; Sumner, Ruge-Jones, and Alcorn 2018) that aid the creation and exponential multiplication of 'digital bullets' – in forms of tweets and perceived detestable comments – that impend over the defense lines of political regimes.

## 3.  METHODS

A qualitative approach informed by an interpretivist epistemological position is adopted by this study, through the conduct of eleven (11) key informant interviews with social media users and influencers in Nigeria. The interviews were conducted between February and April 2020. Twitter users and influencers were engaged for this study, Twitter being a notable platform through which the public engages political discourses in Nigeria (Opeibi, 2019; Kperogi, 2016). Key concepts and gaps that emerged from the literature review informed the questions that guide this study. A purposive sampling approach was used to select respondents, in a bid to draw in depth information from consistent Twitter users, who were easily identified by their following and Twitter engagement metrics. Patterns and recurrent themes that emerged from the transcribed interviews served as points of analysis discussed within the framework of Acemoglu and Robinson's (2008) conceptualization of *de-jure* and *de-facto* political power. Data was also drawn from secondary sources - newspaper publications, government reports and web posts, to corroborate findings from the interviews conducted. This qualitative research is limited, first, by the number of interviews conducted which reduces the extent to which the findings of this research are generalizable. Also, respondents did not include government representatives, which creates a gap in the knowledge that should be drawn about perceptions of the Nigerian government on specific features and affordances of social media platforms, and how, if at all, they stir their repressive actions. However, these limitations do not discount the findings of this study. It is hoped that future research would expand on the objectives – putting into consideration the gaps – of this research towards arriving at more in deep and generalisable findings.

## 4. SOCIAL MEDIA AND LEGISLATIONS IN NIGERIA

Nigeria has repetitively sought to join a league of African countries in taking legal and repressive actions in censoring social media contents which its finds ignoble. As captured by Vareba et al (2018),

> Nigeria has adopted various tools to censor the cyberspace within its territory. While some of these tools have been described as natural, others have virtually been extra judicial and quasi-obnoxious in nature. The country's efforts towards internet censorship can be seen in the bills adopted by the country to criminalize cybercrimes as well as in the various draconian and antidemocratic bills adopted circumstantially by the government to deal with cases of (perceived) abuses of the Internet by political adversaries.

Since the incumbent government of President Muhammadu Buhari assumed power in 2015, bills against the use of social media have been severally rebranded and presented at the floor of the Nigerian Senate. First, in 2015 was the 'Frivolous Petitions and Other Matters Connected Therewith Bill' which was proposed by a Senator in Nigeria's 9$^{th}$ assembly. The bill which targeted online media platforms such as Facebook, Twitter, Instagram, YouTube, WhatsApp and online blogs was widely criticized by Nigerians, describing it as an infringement on individual rights and free speech. However, following public pressure and the report of the Senate Committee on Judiciary, Human Rights and Legal Matters, this bill was withdrawn in the Nigerian Senate on Tuesday May 17, 2016 (Eribake, 2016). This was followed by a sister bill, resuscitated in the Nigerian Senate sooner than was expected, in 2018. This was equally claimed to have been targeted at regulating the use of social media in spreading hate speeches, fake news and false accusations. Also, in preparation for the concluded 2019 general elections, the bill was described to be a measure to curtail the activities of unscrupulous elements who might through the aid of the social media seek to bring down perceived political enemies (Opusunju, 2018). Although the bill passed first reading in the Nigerian Senate, it made no significant headway afterwards, as Nigerians did not fail to mobilize actions and voices against the bill.

Most recently was the re-introduction of a bill titled the 'Protection from Internet Falsehood and Manipulations Bill, 2019', by the Nigerian legislature. The bill which was targeted at penalizing false statements communicated to one or more end-users in Nigeria through the internet and on social media platforms, among other reasons, sought to prevent the transmission of false statements/declaration of facts in Nigeria and to enable measures to be taken to counter the effects of such transmission. Although the sponsor of the bill, a senator representing Niger-East Senatorial district was accused of plagiarizing a similar law in Singapore, passed by the Singaporean Parliament on May 8, 2019 and assented to by the President on June 3, 2019. The senator however defended the claims. According to his tweet on November 23, 2019, "it is therefore inevitable that lessons be drawn from other jurisdictions in fashioning out workable solutions in our own country" (Aborisade, 2019).

The specific objectives of the bill are:

- To prevent the transmission of false statements or declaration of facts in Nigeria
- To end the financing of online mediums that transmit false statements
- To detect and control inauthentic behavior and misuse of online accounts

- When paid content is posted towards a political end, there will be measures to ensure the poster discloses such information
- There will be sanction for offenders (Iroanusi, 2019).

Angry demonstrations by Nigerians trailed the introduction of the bill calling for it to be scrapped. Critics also described the bill as capable of gaging free speech. Further reaction against the bill was seen in a petition that was signed by tens of thousands of Nigerians, unanimously rejecting the bill. Twitter influencers were also not silent about the bill, as their various tweets and posts generated comments and engagements in cumulative millions, which put the government under pressure. A Twitter voice in Nigeria, when interviewed, believes the bill is nothing but:

> …an infringement on the constitutional freedom of speech and will not but deepen public intimidation which the government has always desired.

Another, as well, corroborates this point by affirming that:

> if the government should succeed with the bill, then the country must be ready for a level of disrespect of the rule of law that it has never witnessed before.

A respondent believes that "the freedom of speech is all that the country has left, should it be taken from us, forget it". She goes further to praise the freedom of speech that the country enjoyed under preceding administrations in Nigeria. A close reference by a respondent was a tag on the bill as "… war against our collective freedom". Another respondent appears curious about why the bill has consistently resurfaced in rebranded forms since 2015, the resolution was, however, that there must be some scrupulous machinations going on behind the scenes by some agents of the government.

A contrary viewpoint to the above expressed by an activist who described herself as politically neutral was that:

> This is democracy, the will of the majority will surely prevail, the bill must be allowed to undergo due process. Some are against the bill today cos they know their ways have always been bad. They should change and stop spreading fake news.

Some respondents acknowledge elements of government control via the bill. A student activist in South-West Nigeria said:

> The government is aware about how powerful the voices of the Nigerian Youths are. One good thing about the social media is how information spreads so fast and easily. This is what the government hates. They don't want us to announce their ills to the world. They can't stop us, we refuse to be prisoners in our own land.

Another student activist in South-West Nigeria finds the system of government in Nigeria to be an "unhealthy mixture of democracy and dictatorship, and this is even seen in our universities. There is war against the power and will of the people". References were further made to Nigeria's youngest Senator whose news (though leaked CCTV footage) a few weeks before the proposition of the social media bill, went viral on social media. He was captured by the CCTV of a sex toy shop assaulting a young woman. The senator was recorded justifying the need for the bill, when it was presented at the senate. A few respondents found that his support of the bill may be attributed to the Senator's recent viral condemnation on social media. A respondent suggested that without the social media "these barbaric acts by politicians will go unnoticed". Another claims he would accept the justifications presented by the senator in support of the bill, provided same penalties as spelt by the bill would be defined for corrupt practices by politicians.

Nigerians continued to express their concerns about the bill from November 2019 when it was first proposed in the National Assembly till March 2020 when a public hearing was held, – an event that shredded the prospects of the bill. Although, at the current time of writing, there are indications that Nigerians would have to deal with more of this in the future.

## 5.  DISCUSSION

The interaction between *dejure* political power (as determined by political institutions – democracy in this case) and *de facto* political power (as determined by the organisation of different groups), serves as bedrock to economic and development outcomes. Acemoglu and Robinson (2008) clarifies that although *de facto* political power drives the determination and distribution of economic resources, these resources are not often allocated by institutions, they are rather possessed by groups on the basis of their wealth or their ability to solve their problems of collective action. Evidence in this article presents the social media has a platform for organisation, mobilisation and claim-making (Dawson, 2012), which can be leveraged for the reinforcement of their *de jure* political power towards securing the institutions requisite for development outcomes. This primarily spotlights the social media as a target for repression, given its privileges for collective demand and claim-making.

The bill emphasized 'false declaration', and 'false statements'. By false statements, it captured statements that are likely to: be prejudicial to the country's security, public health, public safety, public tranquility or finances; influence the outcome of an election or referendum; and incite feelings of enmity, hatred, among others. The government's concern about these presupposes the vulnerability of their defenses. Government actions can hardly hamper the popularity and spread of messaging on social media (Hofheinz, 2011). Through (re)tweets, (re)grams and comments, users are able to easily create online trends, grams and feeds which most times go viral before the reaction of government censors. By doing these, *inter alia*, social media users can subvert censorship and state-controlled media (Jones, 2017).

Hence, elected representatives are not inaccessible for engagement and perceived virtual attacks. This viewpoint is well condensed by Omojuwa (2019),

> ...the closest thing to engaging a government official online was being able to tweet at an aide who was himself an aide to another aide who was an assistant to the president... What has changed here is that expectations are now higher, the president is your Twitter neighbor. We now have a voice; our voices are being heard….

Netizens have their ways around the defenses of political regimes towards getting their messaging to being heard in closed and innermost spaces of decision making. Social media is, thus, conceived as an embodiment of virtual bullets that impend over the defense lines of political regimes. Consequently, they respond with repression, as metaphorical 'flak jackets'.

Social media comments often signal one of three things, (1) an applaud of good deeds, (2) an expression of indifference or neutrality, or (3) an ardent rejection or criticism of a public action. In the case of the latter, distressed citizens will choose to construct their messaging in ways that reflect the depth of their deprecation, not minding how it is received or perceived by the targets. The people can be said to have a shared understanding of the transitory nature of political power (Acemoglu and Robinson, 2000) – the understanding that they might possess the power to take some actions today but might not wake to same the next morning –, hence, they try to "lock in the political power they have today by challenging political institutions" (Acemoglu and Robinson, 2000). Whichever side of the divide that possesses more political power determines who captures the offerings of economic institutions. Further, political institutions that allocate *de jure* and *de facto* power determines political power itself (Acemoglu and Robinson, 2008). Consequently, netizens join in the contest for this power against elites who, being small in number, and as a result of their control of politics, have more comparative advantage in investing in *de facto* power (Olson 1965). In other words, it is a contest between parties in the quest for political power.

Although, some supported the provisions of the social media bill, finding it necessary to sieve rebellious contents from the social media space of Nigeria. However, most found the bill to portend an unguarded future for the citizens of Nigeria, the country's democratic posture and deference for the rule of law, as provided in the Nigerian constitution (section 22 and 39) – the freedom of expression. Noteworthy is the fact that such perceived obnoxious contents about government policies, actions and inactions are not particularly fruits of hatred or disdain against the government, very importantly, they evince veracities from which the government is being unwittingly veiled. In order to invite government attention and as a form of political participation, the social media is leveraged as a modifier of political attitudes and involvement of users (Rajalakshmi & Velavuthum, 2014), which translates to an expression of political power against regimes (Liqui, 2007; Boker & Elstub, 2015).

Acemoglu and Robinson (2000) connects the diffused participation of non-elite citizens with the constraining of leaders in authoritarian contexts, through protests, revolutions and threats. Although not an authoritarian setting, social media in Nigeria aids diffused participation and mass mobilization for public causes, through demonstrations and campaigns. For instance, campaigns and movements such as the #NotTooYoungToRun, #DGTrends, #BringBackOurGirls, #OccupyNigeria, #RevolutionNow, and most recently, the #ENDSARS protest, among others, are illustrative. The expression of *de facto* political power through these movements do not only help redirect the focus of the government to the concerns of the people, but they also lay foundation for transformative outcomes. As opined by Jung (2016), there is hardly a political regime that can defeat a people who are a united voice on an issue, a voice that is sustained over time and not weakly tied.

The situation in Nigeria cannot be divorced from mis-governance, corruption and leadership impasse. Further to this, governance in Nigeria, legislative preference, law adjudication and status of citizens most recently queries the defining tenets of the institution of democracy and more importantly threatens the expression of political power by the citizens. Patriots will not ignore approximate costless platforms through which the government can be engaged, as means of expressing and securing their political power and eventually the dividends of governance.

## 6. CONCLUSION

This study has discussed the rebranding of social media legislation in Nigeria. Interviews conducted also granted insight into the perceptions of Nigerian netizens about the country's democratic trajectory and what they are up against – the defense lines of the political regime. In Nigeria, it can be argued that the country's gradually widening democratic space and progress made with internet freedom is attributable to the gridlocks that the current political regime has faced at each episode of its social media censorship agenda, since 2015. The concessions are not credible, neither is repression of social media overly attractive, as evinced by the expression of political power by Nigerian netizens (Acemoglu and Robinson, 2000). This study has argued that political power acquired through political institutions (democracy) provide the latitude for demand for changes and policies that are requisite for development in any society. With political power, netizens are able to defeat the defense lines of political regimes, and through the affordances of social media – digital bullets against which regimes seek to protect themselves through 'flak jackets' – censorship.

Notably, these are critical times when Nigeria deserves the collaborative exertion of all sections of governance to alleviate threatening challenges of the citizens. The legislative institution rather than repeatedly rebranding a bill that has not only been greeted by rejections from Nigerians locally and in diaspora, suffered multilateral defeat (as the institution received an ECOWAS court judgement criminalizing the bill), it should take more proactive legislative actions on the demands of its citizens, to which their political power is constantly expressed.

To check media contents against false reports, the government should outlay enlightenment measures and sensitizations through relevant organizations and groups on how to verify stories, against extensive spread, and unhealthy public incitement, rather than sponsor a bill that has been declared needless by existing laws, such as the Cybercrime (prohibition) Act. The hard-earned freedom of a diverse country like Nigeria should not be circumscribed by guised dictatorial gimmicks. While the government and all executors in the echelons of power will continually commit themselves to development initiatives across all sectors, the people who have delegated their trust of leadership to them should be guaranteed freedom of expression and right to express their political power.

## REFERENCES

Acemoglu, D., & Robinson, J. A. (2000). Why Did the West Extend the Franchise? Democracy, Inequality, and Growth in Historical Perspective. The Quarterly Journal of Economics, 115(4), 1167–1199.

Acemoglu, D., & Robinson, J. A. (2005). Economic Origins of Dictatorship and Democracy. Cambridge University Press. https://doi.org/10.1017/CBO9780511510809.

Acemoglu, D., & Robinson, J. A. (2008, June 19). Paths of Economic and Political Development. The Oxford Handbook of Political Economy. https://doi.org/10.1093/oxfordhb/9780199548477.003.0037.

Aborisade, S. (2019, November 24). Anti-social media bill: Senator defends alleged plagiarism of Singapore statute. Punch News. Retrieved from https://punchng.com/anti-social media-bill-senator-defends-alleged-plagiarism-of-singapore-statute/.

Allmann, K. (2016). Everyday a revolution: Mobility, technology, and resistance after Egypt's Arab Spring [Ph.D., University of Oxford (United Kingdom)].https://search.proquest.com/docview/2342013714?pq-origsite=primo.

Amedie, J. (2015). The impact of social media on society. Advanced Writing: Pop Culture Intersections 2. http://scholarcommons.scu.edu/engl_176/2.

Beauvais, E. (2017). Discursive Equality. PhD thesis, University of British Columbia. https://open.library.ubc.ca/cIRcle/collections/24/items/1.0354400.

Besley, T., & Kudamatsu, M. (2006). Health and Democracy. The American Economic Review, 96(2), 313–318.

Bode, L., Vraga, E., Borah, P., & Shah, V. (2014). A new space for political behavior: Political social networking and its democratic consequences. Journal of Computer-Mediated Communication.

Böker, M., & Elstub, S. (2015). The possibility of critical mini-publics: Realpolitik and normative cycles in democratic theory. Representation, 51: 125–44.

Boulianne, S. (2019). Revolution in the making? Social media effects across the globe. Information, Communication & Society, 22(1), 39–54. https://doi.org/10.1080/1369118X.2017.1353641.

Bryer, T., & Zavattaro, S. (2011). Social media and public administration: Theoretical dimensions and introduction to symposium. Administrative Theory & Praxis, 33(3), 325-340.

Cheeseman, N., Fisher, J., Hassan, I., & Hitchen, J. (2020). Social Media Disruption: Nigeria's WhatsApp Politics. Journal of Democracy, 31(3), 145–159. https://doi.org/10.1353/jod.2020.0037

Dawson, M. C. (2012). Protest, performance and politics: The use of 'nano-media' in social movement activism in South Africa. Research in Drama Education: The Journal of Applied Theatre and Performance, 17(3), 321–345. https://doi.org/10.1080/13569783.2012.694028.

Eribake, A. (2016, May 17). Senate withdraws anti-social media bill. Vanguard News. Retrieved from https://www.vanguardngr.com/2016/05/senate-withdraws-frivolous-petitions-bill/amp/.

Evans, L. (2010). Social media marketing: Strategies for engaging in Facebook, Twitter and other Social Media. Que.

Famuyiwa, D. (2019, November 30). Aisha Buhari backs social media bill. Pulse News. Retrieved from https://www.pulse.ng/news/local/aisha-buhari-backs-social-media-bill/crb2ejs.

Fung, A. (2003). Survey article: Recipes for public spheres: Eight institutional design choices and their consequences. Journal of Political Philosophy, 11: 338–67.

Gerlitz, C., & Helmond, A. (2013). The like economy: Social buttons and the data-intensive web. New Media & Society, 15(8), 1348–1365. https://doi.org/10.1177/1461444812472322.

Goodin, R. E. (2007). Enfranchising all affected interests, and its alternatives. Philosophy and Public Affairs, 40: 40–68.

Gumede, W. (2010), 'Building a Democratic Political Culture', in W. Gumede and L. Dikeni (eds), The Poverty of Ideas: South African Democracy and the Retreat of the Intellectuals, Cape Town: Jacana Publishers, pp. 1–34.

Gumede, W. (2016). Rise in censorship of the Internet and social media in Africa. Journal of African Media Studies, 8(3), 413–421. https://doi.org/10.1386/jams.8.3.413_7.

Hofheinz, A. (2011). Nextopia? Beyond revolution 2.0. International Journal of Communication, 5, 1417-1434.

Holt, K., Shehata, A., Stromback, J., & Ljungberg, E. (2013). Age and the effects of newsmedia attention and social media use on political interest and participation: Do social media function as leveller? European Journal of Communication, 28, 19–34.

Hsieh, Y., & Li, M. (2014). Online political participation, civic talk, and media multiplexity: How Taiwanese citizens express political opinions on the Web. Information, Communication & Society, 17, 26–44.

Iroanusi, Q. (2019, November 24). Explainer: Important things to know about Nigeria's proposed social media bill. Premium Times. Retrieved from https://www.premiumtimesng.com/news/top-news/364900-explainer-important-things-to-know-about-nigerias-proposed-social-media-bill.html

Jha, V. & Bhardwaj, R. (2012). The new marketing renaissance: Paradigm shift in social networks. International Journal of Engineering and management Sciences, 3(3): pp 384-387. Retrieved from www.scienceandnature.org.

Jones, M. (2017). Satire, social media and revolutionary cultural production in the Bahrain uprising: From utopian fiction to political satire. Communication and the Public, Sage Journals, 2(2).

Jung, J.-Y. (2016). Social Media, Global Communications, and the Arab Spring: Cross-Level and Cross Media Story Flows. https://doi.org/10.1057/978-1-137-58141-9_2.

Kellam, M., & Stein, E. A. (2016). Silencing Critics: Why and How Presidents Restrict Media Freedom in Democracies. Comparative Political Studies, 49(1), 36–77. https://doi.org/10.1177/0010414015592644.

Kperogi, F. (2016). Networked social journalism: Media, citizen engagement and democracy in Nigeria. In: Mutsvairo, B. (eds) Participatory Politics and Citizen Journalism in a Networked Africa. Palgrave Macmillan, London.

Liqui, L. (2018). The cost of humour: Political satire on social media and censorship in China. Global and Media Communication, SAGE Journals.

Mausolf, J. G. (2017). Occupy the government: Analyzing presidential and congressional discursive response to movement repression. Social Science Research, 67, 91–114. https://doi.org/10.1016/j.ssresearch.2017.07.001.

McQuail, D. (2005). McQuail's mass communication theory (3rd ed.). London: SAGE Publications.

Mueller, D. C. (2007). Torsten Persson and Guido Tabellini, The Economic Effects of Constitutions. Constitutional Political Economy, 18(1), 63–68. https://doi.org/10.1007/s10602-006-9013-x

Newman, N., Dutton, W., & Blank, G. (2014). Social Media and the News (pp. 135–148). https://doi.org/10.1093/acprof:oso/9780199661992.003.0009.

Nyabola, N. (2018). Digital Democracy, Analogue Politics: How the Internet Era is Transforming Politics in Kenya. Zed Books. http://ebookcentral.proquest.com/lib/londonschoolecons/detail.action?docID=5567836

Okocha, D., & Kumar, Y. (2018). Influence of Media Owners' Political Affiliations on Journalistic Professionalism in Ghana.

Olson, Mancur C. (1965) The Logic of Collective Action, Public Goods and the Theory of Groups Cam bridge, MA Harvard University Press.

Omojuwa, J. (2019). Digital: The New Code of wealth. Lagos: A'Lime Media Limited.

Opeibi, T. (2019). The twittersphere as political engagement space: A study of social media usage in election campaigns in Nigeria. Digital Studies/Le champ numerique 9(1): 6, pp. 1-32

Opusunju, O. (2018, June 27). Bill to regulate social media use by Nigerians passes first reading. ITEdge News. http://itedgenews.ng/2018/06/27/bill-regulate-social-media-use-nigerians-passes-first-reading/.

Rajalakshmi, K. & Velayutham, C. (2014). 'Creating' political awareness through social networking – An empirical study with special reference to Tamil Nadu elections, 2011. Journal of Social Media Studies, 1(1), pp. 71–81.

Smeltzer, S., & Keddy, D. (2010). Won't you be my (political) friend? The changing face (book) of socio-political contestation in Malaysia. Canadian Journal of Development Studies/Revue canadienne d'études du développement, 30(3-4), 421-440.

Sumner, Erin M., Luisa Ruge-Jones, and Davis Alcorn. 2018. 'A Functional Approach to the Facebook Like Button: An Exploration of Meaning, Interpersonal Functionality, and Potential Alternative Response Buttons'. New Media & Society 20(4):1451–69. doi:10.1177/1461444817697917.

Tang, G., & Lee, F. F. (2013). Facebook use and political participation: The impact of exposure to shared political information, connections with public political actors, and network structural heterogeneity. Social Science Computer Review, 31, 763–773.

Tarrow, S. G. (2011). Power in movement social movements and contentious politics (3rd ed.). University Press. http://dx.doi.org/10.1017/CBO9780511973529.

The Guardian Editorial Board (2015, December 15). The 'anti-social media' bill. Retrieved from https://m.guardian.ng/opinion/the-anti-social-media-bill/.

Thompson, E.P. (1975). Whigs and hunters: The origin of the Black Act - London School of Economics and Political Science. Retrieved 14 May 2021, from https://librarysearch.lse.ac.uk/primo-explore/fulldisplay/44LSE_ALMA_DS21106348480002021/44LSE_VU1.

Udanor, C., & Anyanwu, C. (2019). Combating the challenges of social media hate speech in a polarized society: A Twitter ego lexalytics approach. Data Technologies and Applications, ahead-of-print. https://doi.org/10.1108/DTA-01-2019-0007.

Van Dijk, J.A.G.M. (2006). Network Society: Social aspects of new media (2nd ed.). London: SAGE Publications.

Vasi, I. B., Walker, E. T., Johnson, J. S., & Tan, H. F. (2015). 'No Fracking Way!' Documentary Film, Discursive Opportunity, and Local Opposition against Hydraulic Fracturing in the United States, 2010 to 2013. American Sociological Review, 80(5), 934–959.

Vareba, A., Nwinaene, V., & Theophilus, S. (2018). Internet censorship and freedom of expression in Nigeria. International Journal of Media, Journalism and Mass Communications (IJMJMC), 3(2), pp. 25-30

Warren, M. E. (2017). A Problem-Based Approach to Democratic Theory. American Political Science Review, 111: 39–53.

Young, I. M. (2000). Inclusion and Democracy. Oxford: Oxford University Press